\documentclass[fleqn,11pt]{article}
\usepackage{amsmath,amssymb,latexsym}
\usepackage{graphicx}

\usepackage{amsfonts,amssymb,cite}

\def\noi{\noindent}

\newcommand{\Title}[1]{\noi {{\Large\bf #1}}\\[1ex]}

\def\Aunames#1{\noi{\bf #1}}
\def\auth#1{${}^{#1}$}
\def\Addresses#1{\medskip\noi \protect
	\begin{description}\itemsep -3pt {\it #1} \end{description}}
\def\addr#1#2{\item[${}^{#1}$]{\it #2}}

\newcommand{\Abstract}[1]{\vskip 2mm \begin{center}
        \parbox{16.4cm}{\small\noi #1} \end{center}\medskip}

\def\email#1#2{\footnotetext[#1]{e-mail: #2}\addtocounter{footnote}{1}}

\def\nq{\hspace*{-1em}}
\def\nqq{\hspace*{-2em}}
\def\nhq{\hspace*{-0.5em}}

\def\cm{\hspace*{1cm}}
\def\inch{\hspace*{1in}}
\def\wide{\mbox{$\dst\vphantom{\int}$}}

\def\Jl#1#2{#1 {\bf #2},\ }

\def\ApJ#1 {\Jl{Astroph. J.}{#1}}
\def\CQG#1 {\Jl{Class. Quantum Grav.}{#1}}
\def\DAN#1 {\Jl{Dokl. AN SSSR}{#1}}
\def\GC#1 {\Jl{Grav. Cosmol.}{#1}}
\def\GRG#1 {\Jl{Gen. Rel. Grav.}{#1}}
\def\JETF#1 {\Jl{Zh. Eksp. Teor. Fiz.}{#1}}
\def\JETP#1 {\Jl{Sov. Phys. JETP}{#1}}
\def\JHEP#1 {\Jl{JHEP}{#1}}
\def\JMP#1 {\Jl{J. Math. Phys.}{#1}}
\def\NPB#1 {\Jl{Nucl. Phys. B}{#1}}
\def\NP#1 {\Jl{Nucl. Phys.}{#1}}
\def\PLA#1 {\Jl{Phys. Lett. A}{#1}}
\def\PLB#1 {\Jl{Phys. Lett. B}{#1}}
\def\PRD#1 {\Jl{Phys. Rev. D}{#1}}
\def\PRL#1 {\Jl{Phys. Rev. Lett.}{#1}}

\def\al{&\nhq}
\def\lal{&&\nqq {}}
\def\eq{Eq.\,}
\def\eqs{Eqs.\,}
\def\beq{\begin{equation}}
\def\eeq{\end{equation}}
\def\bear{\begin{eqnarray}}
\def\bearr{\begin{eqnarray} \lal}
\def\ear{\end{eqnarray}}
\def\earn{\nonumber \end{eqnarray}}

\def\nnn{\nonumber\\ \lal }

\def\yyy{\\[5pt] \lal }
\def\eql{\al =\al}

\def\dst{\displaystyle}
\def\tst{\textstyle}

\def\fract#1#2{{\tst\frac{#1}{#2}}}

\def\half{{\fract{1}{2}}}

\def\e{{\,\rm e}}
\def\d{\partial}

\def\sign{\mathop{\rm sign}\nolimits}

\def\const{{\rm const}}
\def\eps{\varepsilon}

\def\then{\ \Rightarrow\ }

\newcommand{\vars}[1]{\left\{\begin{array}{ll}#1\end{array}\right.}

\def\qua{\fract 14}

\def\wt{\widetilde}

\def\tT{{\wt T}}

\def\M{{\mathbb M}}
\def\R{{\mathbb R}}

\def\asflat{asymptotically flat}

\def\AdS{anti-de Sitter}
\def\GR{general relativity}
\def\wh{wormhole}
\def\whs{wormholes}
\def\sph{spherically symmetric}
\def\ssph{static, spherically symmetric}
\def\EE{Einstein equation}

        \def\kappa{\varkappa}
\def\ua{{\underline a}}      \def\um{{\underline m}}
\def\ok{{\overline k}}

\def\rf{\eqref}
\def\eqn{\eq\eqref}
\def\fin{{\rm fin}}

\begin{document}
\twocolumn[
\vspace{1cm}

\Title{Wormholes leading to extra dimensions}

\Aunames{K. A. Bronnikov\auth{a,b,c,1} and M. V. Skvortsova\auth{b,2}}
   
\Addresses{\small
\addr a {VNIIMS, Ozyornaya ul. 46, Moscow 119361, Russia}
\addr b	{Peoples' Friendship University of Russia,
	ul. Miklukho-Maklaya 6, Moscow 117198, Russia}
\addr c	{National Research Nuclear University ``MEPhI''
	(Moscow Engineering Physics Institute), Moscow, Russia}
}

\bigskip

\Abstract
     {In 6D general relativity with a scalar field as a source of gravity, a new type of static
       wormhole solutions is presented: such wormholes connect our universe with a small 2D extra
       subspace with a universe where this extra subspace is large, and the whole 
       space-time is effectively 6-dimensional. We consider manifolds with the structure
       $\M_0 \times \M_1 \times \M_2$, where $\M_0$ is 2D Lorentzian space-time 
       while each of $\M_{1,2}$ can be a 2-sphere or a 2-torus. After selecting possible asymptotic
       behaviors of the metric functions compatible with the field equations, we give two explicit 
       examples of wormhole solutions with spherical symmetry in our space-time and toroidal extra
       dimensions. In one example, with a massless scalar field (it is a special case of a well-known 
       more general solution), the extra dimensions have a large constant size at the ``far end''; 
       the other example contains a nonzero potential $V(\phi)$ which provides a 6D anti-de Sitter
       asymptotic, where all spatial dimensions are infinite.}       	 

\medskip
] 
\email 1 {kb20@yandex.ru}
\email 2 {milenas577@mail.ru}

\section{Introduction}

  Multidimensional theories suggest a great variety of geometries,  topologies and compactification
  schemes (for reviews see, e.g.,  [1--4] and references therein). However, 
  there still emerge new ideas on how our 4D space-time may be inscribed in a multidimensional 
  world. One recently suggested idea \cite{Rub16} is that of exchanging roles of different 
  dimensions in different parts of space-time. 

  More specifically, considered was \cite{Rub16} a 6D static
  space-time of the form  $\M = \M_0 \times \M_1 \times \M_2$, where $\M_0$ is 2D Lorentzian 
  space-time parametrized by time $t$ and a ``radial'' coordinate $x$ while $\M_{1,2}$ are 
  two-spheres with radii $r_1(x)$ and $r_2(x)$. Moreover, as $x \to -\infty$, $r_1 \to \infty$ 
  (so that the 4D space-time $\M_0 \times \M_1$ is \asflat) while $r_2$ tends to a small constant
  value, and $\M_2$ is thus a spherical extra space. The same picture occurs in the other 
  asymptotic region, $x \to +\infty$, but, on the contrary, $\M_1$ is small and $\M_2$ is large, and now
  $\M_0 \times \M_2$  is \asflat. The spatial section of $\M_0 \times \M_1$ looks like a funnel open to
  the left and narrow on the right, that of  $\M_0 \times \M_2$ like a funnel open to the right and narrow 
  on the left. The whole structure resembles a \wh\ (and we suggest to call it a {\it Rubin \wh}) 
  since it connects two large space-time regions, but now these regions belong to different sections of
  a multidimensional manifold.\footnote
	{For reviews of \wh\ physics in different contexts, in particular, in theories of gravity 
	alternative to GR see, e.g., \cite{BR, visser-book, lobo} and also more recent 
         papers, e.g.,  \cite{dzhun, almaz, sunny}.}   
   
  Such a solution was obtained in \cite{Rub16} in a certain approximation from $R^2$ gravity. 
  Inspired by this work, we tried to find similar solutions in 6D \GR\ (GR) with a minimally coupled 
  scalar field as a source. Such an attempt looks natural since $f(R)$ gravity is known to be 
  equivalent to a certain class of scalar-tensor theories whose Einstein frame formulation 
  has the form of GR with a a minimally coupled scalar field.
  For more generality, we admit both spherical and toroidal forms of the compact 2D manifolds 
  $\M_1$ and $\M_2$. Toroidal geometry of extra dimensions is often considered, 
  and it should be noted that this symmetry in our universe also seems to be compatible  
  with observations \cite{stei1, stei2}.

  It turns out, however, that in this framework Rubin \wh\ solutions do not exist since the field 
  equations do not admit the needed asymptotic behavior, but, instead, other 
  structures of interest are discovered: these are \whs\ which connect an effectively 
  4D space-time region where extra dimensions are small with an effectively multidimensional 
  region, where the potentially observable physical picture should be drastically different from ours. 
  We give two explicit examples of such solutions, one with a zero potential (actually, a special 
  case of a general solution known for a long time), the other with a nonzero potential; in both cases,
  the scalar field should be phantom, i.e., have a wrong sign of kinetic energy.  The necessity of 
  a phantom (or exotic) nature of a source of gravity for obtaining \wh\ and other regular models 
  in \GR\ and its many extensions like scalar-tensor and $f(R)$ gravity is well known 
  (see, e.g., \cite{BR, hohvis, bsta1, bsta2}) and can be avoided by invoking alternative 
  geometries (e.g., \cite{BKim, almaz}) or/and more general gravitational actions 
  \cite{GB1, GB2}.  There are theoretical arguments both {\it pro et contra} the possible 
  existence of phantom fields, see, e.g., discussions in 
  \cite{bsta1, BFab07, visser-book, lobo}. In this paper, as in many 
  others, we admit it as a working hypothesis.

  The next section presents the field equations for the system to be studied. Section 3 is 
  devoted to an analysis of asymptotic properties of the metric admitted by the field equations.
  Section 4 describes two examples of \wh\ models with large extra dimensions beyond the throat,
  and Section 5 contains our concluding remarks.   
  
\section{Equations in 4+2 dimensions}

  We consider 6D GR with a minimally coupled scalar field $\phi$ with a potential $V(\phi)$
   as the only source of gravity. So the total action is 
\beq    \nq         \label{act}
             S = \frac{m_6^2}{2} \int \sqrt{|g_6|}
	 \Big[R_6 + 2\eps_\phi g^{AB} \d_A\phi \d_B\phi - 2V(\phi)\Big],
\eeq
  where $m_6$ is the 6D Planck mass, $R_6$ and $g_6$ are the 6D Ricci scalar and metric
  determinant, respectively, $\eps_\phi = 1$ for a normal, canonical scalar field, $\eps_\phi =-1$ for 
  a phantom one, and $A, B, \ldots = \overline{0,5}$.  The corresponding equations 
  of motion are the scalar field equation $2\eps_\phi \Box_6\phi + dV/d\phi =0$ and the 
  \EE s which can be written as
\bearr             \label{EE}
             R^A_B = - \tT^A_B \equiv - T^A_B - \qua \delta^A_B T^C_C 
\nnn \cm
                         \equiv - 2\eps_\phi  \d^A\phi \d_B\phi + \half V(\phi) \delta^A_B,
\ear
  where $R^A_B$ is the 6D Ricci tensor and $T^A_B$ is the stress-energy tensor 
  (SET) of the scalar field.

  Now, we consider the 6D manifold with the structure of a direct product of three 2D spaces,
  $\M = \M_0 \times \M_1 \times \M_2$, where $\M_0$ is 2D space-time 
  with the coordinates $x^0 = t$ and $x^1 =u$, while $\M_1$ and 
  $\M_2$ are compact 2D spaces of constant nonnegative curvature, i.e., each of them can
  be a sphere or a torus. The  metric is taken in the form:
\beq                                                 \label{ds_6}
	ds^2 = \e^{2\gamma} dt^2 - \e^{2\alpha}du^2 - \e^{2\beta} d\Omega_1^2
	  - \e^{2\lambda}d\Omega_2^2,
\eeq
  where $\alpha,\beta,\gamma,\lambda$  are functions of an arbitrarily chosen
  ``radial'' coordinate $u$, while $d\Omega_1^2$ and  $d\Omega_2^2$ are $u$-independent metrics 
  on 2D manifolds $\M_1$ and $\M_2$ of unit size. We also assume $\phi = \phi(u)$.    

  So, we do not fix which of $\M_{1,2}$ belongs to our 4D space-time and which is ``extra'':
  everything depends on their size. For example, if $\M_1$ is large and spherical while $\M_2$ 
  is small and toroidal, we have a \ssph\ configuraton in 4D and a toroidal extra space, and so on. 

  The nonzero components of the Ricci tensor are
\bear
	R^t_t \eql -\e^{-2\alpha}\Big[ \gamma_{uu}
		+ \gamma_u \eta_u \Big], 		\label{Rtt}
\\                                          \label{Ruu}
	R^u_u \eql - \e^{-2\alpha}\Big[
	   \gamma_{uu} + \gamma_u^2 - \alpha_u\gamma_u
	 + 2\beta_{uu} + 2\beta_u^2 
\nnn \qquad
	- 2\alpha_u\beta_u 	
            +2\lambda_{uu} + 2\lambda_u^2 - 2\alpha_u\lambda_u\Big], 	   
\\
	R^{\ua}_a \eql \eps_1 \e^{-2\beta} -\e^{-2\alpha}
		\Big[ \beta_{uu} + \beta_u \eta_u\Big],					\label{R22}
\\
	R^{\um}_m \eql  \eps_2 \e^{-2\lambda} -\e^{-2\alpha}
		\Big[ \lambda_{uu} + \lambda_u \eta_u\Big],					\label{R55}
\ear
  where the subscript $u$ denotes $d/du$; 
  the indices $a, b, ...=  2,3$ (they belong to $\M_1$); $m,n,... = 4,5$ (they belong to $\M_2$); 
  there is no summing over an underlined index; $\eps_1 =1\ {\rm or}\ 0$ if $\M_1$ is a sphere or 
  a torus, respectively, and similarly for $\eps_2$; lastly,
\beq                                                        \label{eta6}
	\eta \equiv -\alpha + \gamma +2\beta +2\lambda.
\eeq

  We notice that the scalar field equation is a consequence of the \EE s, and the SET of 
  the scalar field has the components
\bearr                                  \label{sym-T}
              \tT^t_t = \tT^\ua_a = \tT^\um_m = - \half V(\phi), 
\nnn
              \tT^t_t - \tT^u_u = 2\eps_\phi \e^{-2\alpha} \phi_u^2.  
\ear
  
  Let us choose the quasiglobal coordinate $u = x$, such that $\alpha+ \gamma =0$, and denote
\bearr
             \e^{2\gamma} = \e^{-2\alpha} = A(x), 
\nnn
               \e^{2\beta} = r^2(x) \equiv R(x), 
\nnn  
               \e^{2\lambda} = p^2(x) \equiv P(x).
\ear
  Due to the symmetry of the problem and the properties \rf{sym-T} of the SET, 
  there are four independent equations, and it is convenient to use the following ones
  (the prime denotes $d/dx)$):
\bearr     \nhq                             \label{00}
             R^t_t = - \tT^t_t \ \then \ \ - \frac{1}{PR}(A' PR)' = V(\phi),
\yyy  	     \nhq                            \label{01}     
             R^t_t {-} R^x_x = -\tT^t_t {+}\tT^x_x  \  \then \ \
				\frac{r''}{r} + \frac{p''}{p} = -\eps_\phi  \phi'{}^2,
\yyy        \nhq                           \label{02}
             R^t_t {-} R^{\ua}_a = 0  \ \then \ \ [P (AR' - A'R)]' = 2\eps_1 P, 
\yyy         \nhq                           \label{05}
             R^t_t {-} R^{\um}_m = 0 \ \then \ \ [R (AP' - A'P)]' = 2\eps_2 R.
\ear
 
  Equations \rf{02} and \rf{05} contain only the metric functions $A(x), P(x),  R(x)$. 
  Therefore, considering them separately, these are two equations for three unknown 
  functions, so there is arbitrariness in one function; if these functions are known, the other 
  two \EE s can be used to find the scalar $\phi$ and the potential $V$. From \eqn{01} it 
  follows that solutions with $r > 0$ and $p > 0$ in the whole range $x \in \R$ can only exist
  with $\eps_\phi = -1$, i.e., a phantom field, since such solutions require $r''>0$ and $p'' >0$. 

\section {Possible asymptotic behavior of the metric}

  The metric under consideration describes the following types of geometries:

\medskip\noi
  {\bf 1. SS (double spherical) space-times:} the case $\eps_1 = \eps_2 =1$. If the spheres 
  $\M_1$ are large  and $\M_2$ are small (or vice versa), we have static spherical symmetry 
  in our space-time and a spherical extra space. It is also possible that both spheres are large, 
  then we have a 6D space-time where all dimensions are observable.

\medskip\noi
 {\bf  2. ST (spherical-toroidal) space-times:} the case $\eps_1 =1,\ \eps_2 =0$ (or vice versa). 
  If $M_1$ is large and $\M_2$ small, we have static spherical symmetry in our space-time 
  and a toroidal extra  space. The opposite situation is also possible as well as a total observable 
  6D geometry.

\medskip\noi
  {\bf  3. TT (double toroidal) space-times:} if $\eps_1 = \eps_2 =0$, we have the same as 
  before but both $\M_1$ and $\M_2$ are toroidal. 
  
  Our interest is in finding configurations where $x \in \R$ and there are different geometries
  in the two asymptotic regions $x\to \pm\infty$. In particular, there can be two 4D flat 
  asymptotic regions at large positive and negative $x$, where at one end the large 4D space 
  contains $\M_1$, parametrized by the coordinates $t, x, x^a$ ($a = 2,3$),
  while at the other end such a 4D space contains $\M_2$ and is parametrized by $t, x, x^m$ 
  ($m = 4,5$), and in each case the remaining 2D subspace has a small constant size. 
  Other solutions are thinkable, where one or both asymptotic regions have the \AdS\ (AdS)
  geometry. In this section we do not consider the properties of the scalar field but only 
  analyze which  kinds of asymptotic behavior are compatible with \eqs \rf{02} and \rf{05} 
  for each of  the types 1--3 of 6D geometry.  
  
\subsection{Double spherical (SS) space-times} 

  {\bf SS1}. Consider first an \asflat\ 4D space-time with constant extra dimensions.
  Without loss of generality, this means that\footnote
	{Here and henceforth the symbol ``$\fin$'' means a positive constant.}
\beq                                                                                                      \label{flat+}
	A(x) \to \fin, \quad\ R(x) \sim x^2, \quad\ 
                                                        P(x) \to \fin 
\eeq
  as  $x\to \infty$. Let us substitute these conditions to \eqs \rf{02} and \rf{05}. According to \rf{flat+},
  $R' \sim x$, $A' \sim x^{-2}$ or even smaller (due to the expansion 
  $A = A_- + A_{-1}/x + \ldots$), and the l.h.s. of \rf{02} tends, in general, to a nonzero 
  constant, which agrees with the requirement to $P$ that stands on the r.h.s.. However, in 
  \rf{05} the expression in square brackets tends to a constant, hence its derivative 
  vanishes, while the r.h.s., equal to $2R$, should behave as $x^2$. 

  We conclude that {\it the asymptotic conditions \rf{flat+} are incompatible with the field 
  equations.}

  On equal grounds we could consider $x\to -\infty$ and/or exchange $R(x)$ and $P(x)$. 

\medskip\noi
  {\bf SS2}. Next, an asymptotically AdS 4D geometry with constant extra dimensions 
  corresponds to
\beq                                                                                                  \label{AdS+}
	A(x) \sim x^2, \quad\ R(x) \sim x^2, \quad\ 
                                                        P(x) \to \fin
\eeq
  as  $x\to \infty$. Assuming the expansions $A(x) = A_2 x^2 + A_1 x + \ldots$, 
  $R(x) = R_2 x^2 + R_1 x + \ldots$ and substituting them to \eqn{02}, it is easy to see that 
  in square brackets there is $O(x^2)$,
  hence its derivative is $O(x)$ while we need it to be $O(1)$ to satisfy the equation. However, 
  such a behavior is achieved under the condition $A_1 R_2 = A_2 R_1$ for the expansion 
  parameters. Furthermore, in \eqn{05} one has $A P' = O(1)$, $PA' = O(x)$, hence in the whole
  square bracket there is $O(x^3)$ which agrees with $O(x^2)$ on the r.h.s.. However, the
  expression on the l.h.s. is necessarily negative and cannot be equal to 2R. A similar
  conclusion could be obtained by considering the limit $x\to -\infty$. Therefore, as before, 
  {\it the behavior \rf{AdS+} is incompatible with the field equations.}

\medskip\noi
  {\bf SS3}. Let us check whether both spheres $\M_1$ and $\M_2$ can be asymptotically large,
  so that $R \sim x^2$ and $P\sim x^2$ as $x \to \pm \infty$.  
  An inspection similar to the previous one shows that such a behavior can occur both 
  with  $A\to \fin$ and $A\sim x^2$, though in the latter case a solution is only possible under 
  special conditions on the expansion parameters of the functions involved.

\subsection{Spherical-toroidal (ST) space-times} 

  For this kind of geometry,  \eqn{02} has the same form as before, but \eqn{05} has now a 
  zero r.h.s., and its first integral reads											
\beq                         \label{05i}
                A^2 R (P/A)' = K = \const. 
\eeq       
  We have to distinguish the cases $K\ne 0$ and $K=0$.   

\medskip\noi
  {\bf ST1}. Consider the conditions \rf{flat+}. As before, \eqn{02} does not contradict them.
  As to \rf{05i}, in its l.h.s. we have $R\sim x^2$ while, in general, $(P/A)' \sim x^{-2}$,
  therefore \eqn{05i} can hold with $K \ne 0$. If $K=0$, we simply have $P(x) = kA(x)$,
  which is also admissible.

\medskip\noi
  {\bf ST2}. Unlike SS space-times, here the metric coefficients $R$ and $P$ are not 
  equivalent, therefore, besides \rf{flat+}, we should consider the conditions for
  an \asflat\ toroidal 4D space-time with spherical extra dimensions such that
\beq                                                                                                      \label{flat0}
	A(x) \to \fin, \quad\ P(x) \sim x^2, \quad\   R(x) \to \fin
\eeq
   as  $x\to \infty$. Proceeding as before, we see that the l.h.s. of \eqn{02} vanishes at 
   infinity, contrary to a growing r.h.s.. Thus such a behavior is impossible.

\medskip\noi
  {\bf ST3}. Consider an asymptotically AdS spherical 4D space-time with toroidal extra dimensions,
  we return to the conditions \rf{AdS+} and see that \eqn{02} can hold as $x\to\infty$ 
  under the condition $A_1 R_2 = A_2 R_1$, as in item SS2. However, \eqn{05i} cannot 
  hold since its l.h.s. grows as $x^3$. So this behavior is excluded.

\medskip\noi
  {\bf ST4}. The opposite case of an asymptotically AdS toroidal 4D space-time with spherical 
  extra dimensions corresponds to the conditions   
\beq                                                                                                  \label{AdS0}
	A(x) \sim x^2, \quad\ P(x) \sim x^2, \quad\ R(x) \to \fin 
\eeq
  as  $x\to \infty$. In \eqn{02} we obtain the same situation with signs as in item SS2, 
  excluding this kind of behavior.   

\medskip\noi
  {\bf ST5}. In the same way as in item SS3, it can be verified that solutions where 
  both $\M_1$ (sphere) and $\M_2$ 
  (torus) are asymptotically large, are not excluded. so that $R \sim x^2$ and $P\sim x^2$ as 
  $x \to \pm \infty$. An inspection again shows that such a behavior can occur both 
  with  $A\to \fin$ and $A\sim x^2$, but if $K\ne 0$ in \eqn{05i}, there emerge  
  special conditions on the expansion parameters of the functions involved. If $K=0$,
  then \eqn{05i} simply leads to $P(x) = kA(x)$, and solutions where all three functions
  grow as $x^2$ are allowed.   

\subsection{Double toroidal (TT) space-times} 

  In a TT system, in addition to \rf{05i}, we have an integral of \eqn{02}:
\beq                         \label{02i}
                A^2 P (R/A)' = L = \const. 
\eeq       
  Now $R(x)$ and $P(x)$ are again interchangeable, which reduces the number of opportunities.

\medskip\noi  
  {\bf TT1}. Consider the opportunity \rf{flat+}. Then from \rf{02i} it follows either 
  $R \sim x$ or $R \to \fin$, both cases contrary to the assumption $R \sim x^2$. 
  Substituting it to \eqn{05i}, we see that (since at best $(P/A)'\sim 1/x^2$) its l.h.s. vanishes 
  at large $|x|$, which leads to $K =0$, hence $P = cA$, $c = \const$. Thus $R\sim x$ and 
  $P= cA \to \fin$ is a possible asymptotic behavior of our solution. 

\medskip\noi
  {\bf TT2}. The conditions \rf{AdS+}, being substituted to \rf{05i}, leads to a l.h.s. growing as 
  $x^3$, so this behavior is excluded.

\medskip\noi
  {\bf TT3}.  As before, solutions with both $R \sim x^2$ and $P\sim x^2$ as $x \to \pm \infty$
  are not excluded, but, as follows from \rf{05i} and \rf{02i}, only with  $A(x)$ growing in 
  the same manner and only under special conditions on the expansion parameters, leading 
  to $R/A = \const + O(x^{-5})$ and $P/A = \const + O(x^{-5})$.

\medskip\noi
  {\bf TT4}. In the case $K =L = 0$, a ``trivial'' asymptotic behavior where all three functions tend to
  constant values, is also compatible with the equations. One or both sizes $R$ and $P$ can certainly 
  be large to make the corresponding 2-space visible.  

  The results of this analysis are summarized in the table which shows rather 
  a narrow choice of opportunities.

\begin{table}[ht]
\caption{\small Asymptotic behaviors compatible with \eqs \rf{02} and \rf{05}. Notations: the 
           symbols $+$ or $-$ mean that the corresponding behavior is possible or impossible, 
           respectively, $\pm$ that it is possible under special conditions for the functions involved, 
           and n/a that such an opportunity is not applicable. The words ``4D flat spherical'' mean 
           a 4D \asflat\ \sph\ space-time and the size of  the extra subspace tends to a constant, 
          and so on.}
\medskip
\centering
\begin{tabular}{|c|c|c|c|}
\hline
Asymptotic          & \multicolumn{3}{|c|}{6D geometries}  \\
         \cline{2-4}
   behavior         &     SS        &    ST   &     TT    \\ 
\hline
   4D flat spherical    &  --       &   +     &    n/a  \wide \\
   4D flat toroidal      & n/a      &    --    &   $\pm$ \\ 
   4D AdS spherical  &  --       &   --     &    n/a  \wide \\
   4D AdS toroidal    &  n/a     &   --     &    --   \\ 
   6D AdS                &  $\pm$ &$\pm$ &   $\pm$  \wide \\
\hline
\end{tabular}
\end{table}
         
\subsection{Possible solutions}
  
  We see that in SS geometry the only possible asymptotic conditions 
  among those we have considered are those where all dimensions are large.

  In ST geometry, in addition to such effectively 6D asymptotics, there is one more 
  opportunity with \asflat\ \sph\ 4D space-time and constant extra dimensions.   
  Thus, having a usual 4D space-time at one end, we can have another similar asymptotic 
  (though maybe with a drastically different size of extra dimensions),
  or arrive at an effectively 6D space-time with $R(x)$ and $P(x)$ both growing.
  In what follows we will give examples illustrating both opportunities.  

  Similar variants exist in TT geometry: 4D space-time can be \asflat\
  ($R \to \const$ or slowly growing, with $R\sim x$) with constant extra dimensions, 
  or there can be all six large dimensions. Such solutions will not be considered here.    

\section {Examples}

  Of utmost interest for us are space-times with \asflat\ \sph\ geometry in one asymptotic region 
  and something different in the other. As follows from the above-said, there can be two kinds of 
  such solutions from the ST class: (i) wormholes with strongly different size of extra dimensions
  at the two ends, which exist, in particular, among well-known solutions for a massless scalar, 
  $V=0$ \cite{k95, bim97}, and (ii) wormholes with infinitely growing extra dimensions at the 
  other end. 

\subsection {Example 1: ST wormholes with a massless scalar}  

  Consider \eqs \rf{EE} for the ST metric \rf{ds_6} (that is, $\eps_1=1$, $\eps_2 =0$) and a 
  massless ($V=0$) scalar field $\phi$. Following the well-known method 
  \cite{k95, bim97, k73}, we choose the gauge $\eta =0$ (see \rf{eta6})\footnote 
	{This choice of the $u$ coordinate, the ``harmonic'' one \cite{BR, k73}  
           (specified by $\eta \equiv = -\alpha + \gamma +2\beta + 2\lambda = 0$), is quite different 
           from the ``quasiglobal'' one, $u=x$ (specified by $\alpha + \gamma = 0$), used in 
           \eqs \rf{00}--\rf{05} and further in most of  the paper. So, in this subsection,  instead of  
           \eqs \rf{00}--\rf{05}, we directly use the expressions \rf{Rtt}--\rf{R55} with $\eta=0$.}    
  and solve the equation $R^u_u + R^\ua_a =0$,
  which has the form $(\alpha - \beta)_{uu} = \e^{2\alpha-2\beta}$ and gives 
\beq                 \label{s}
	  \e^{\beta-\alpha} = s(k,u) := \vars {               
                    k^{-1} \sinh ku, & k>0,\\
                    u,               & k=0,\\
                    k^{-1}\sin ku,   & k<0,     }
\eeq   
  where $k$ is an integration constant, and one more constant has been removed by 
  choosing the zero point of $u$. Other \EE s lead to $\gamma_{uu} = \lambda_{uu}=0$,
  and the resulting metric reads
\bearr                                     \label{ds_1}
       ds^2 = \e^{2mu}dt^2 - \frac{\e^{-2mu-4nu}}{s^2(k,u)} 
	\Big ( \frac{du^2}{s^2(k,u)} + d\Omega_1^2 \Big)
\nnn \cm
		 - \e^{2nu} d\Omega_2^2,
\ear
  where $m, n$ are integration constants; two more constants are excluded by 
  choosing the time scale and a length unit equal to the size of the toroidal subspace $\M_2$
  at $u=0$, which now corresponds to flat spatial infinity in $\M_0 \times \M_1$.
  Without loss of generality we take $u > 0$ for the whole $u$ range of the solution. 

  Next, the scalar field equation leads to $\phi = Cu$, $C = \const$, and 
  lastly, there is a relation between the constants that follows from the constraint equation
  $R^u_u - \half R_6 = - T^u_u$:
\beq                              \label{int}
           k^2 \sign k = 2\eps_\phi C^2 + m^2 + 3n^2 + 2mn.
\eeq 

  Of interest for us is the solution with $k <0 $, which exists if and only if $\eps_\phi = -1$.  
  In this case the $u$ coordinate ranges from 0 to $\pi/|k|$, it can be easily verified that
  the metric in $\M_0 \times \M_1$ is \asflat\ as $u \to \pi/|k|$, and the whole metric is 
  everywhere regular. It is thus a \wh\ geometry, as required.
   
  Let us consider, for simplicity. the case $m=0$ (a ``force-free'' gravitational field with zero
  Schwarzschild mass since $g_{00} \equiv 1$), denote $-k = \ok > 0$ and make the substitution  
\beq
	\ok u = \cot^{-1} (-z/\ok).	
\eeq
   The metric takes the form 
\bearr                                       \label{ds_2}
       ds^2 = dt^2 - \e^{-4nu}\big[dz^2 + (z^2+ \ok^2) d\Omega_1^2\big]
\nnn \inch
		 - \e^{2nu}d\Omega_2^2.  
\ear
  It describes a \sph, twice \asflat\ wormhole in the 4D subspace $\M_0 \times \M_1$
  with a toroidal extra space $\M_2$ having a unit size, $p_-$ at $u=0$ 
  (that is, $z = -\infty$) and the size $p_+ = \e^{n\pi/\ok} p_-$ at the other end, 
  $u = \pi/\ok$, corresponding to $z = + \infty$.\footnote
	{In the trivial case $n=0$ we obtain the well-known 4D Ellis \wh\ \cite{k73, hell} 
	 times a toroidal extra space of constant size.}

 The \wh\ throat is a minimum of $r(z) = \e^\beta(z) = \e^{-2nu}(z^2 + \ok^2)^{1/2}$, 
 it is located at $z = 2n$ and has the radius
\beq               \label{r_th} 
	r_{\min} = \sqrt{\ok^2 + 4n^2} \exp{\Big(\frac{2n}{\ok} \cot^{-1} \frac{2n}{\ok}\Big)}.
\eeq
   
  Suppose that the size of extra dimensions $p_-$ on the left end, $z=-\infty$, is small 
  enough to be invisible by modern instruments, say, $p_- = 10^{-17}$ cm. It is clear that 
  the size $p_+$ on the other end will be much larger if we take a large enough value of
  $n/\ok$. For example, to obtain $p_+ \sim 1$ m, one should take $n/\ok \approx 14$.
\begin{figure*}
\centering
\includegraphics[width=7cm,height=5cm]{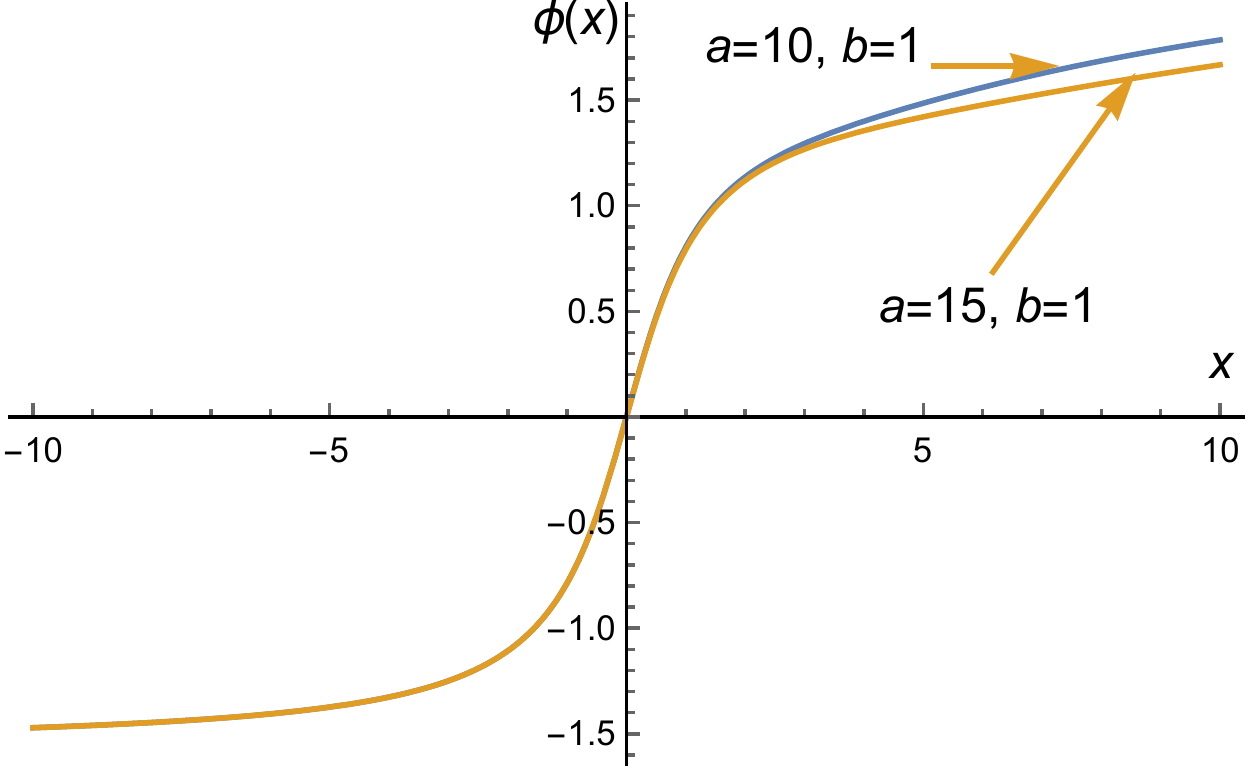}
\qquad
\includegraphics[width=7cm,height=5cm]{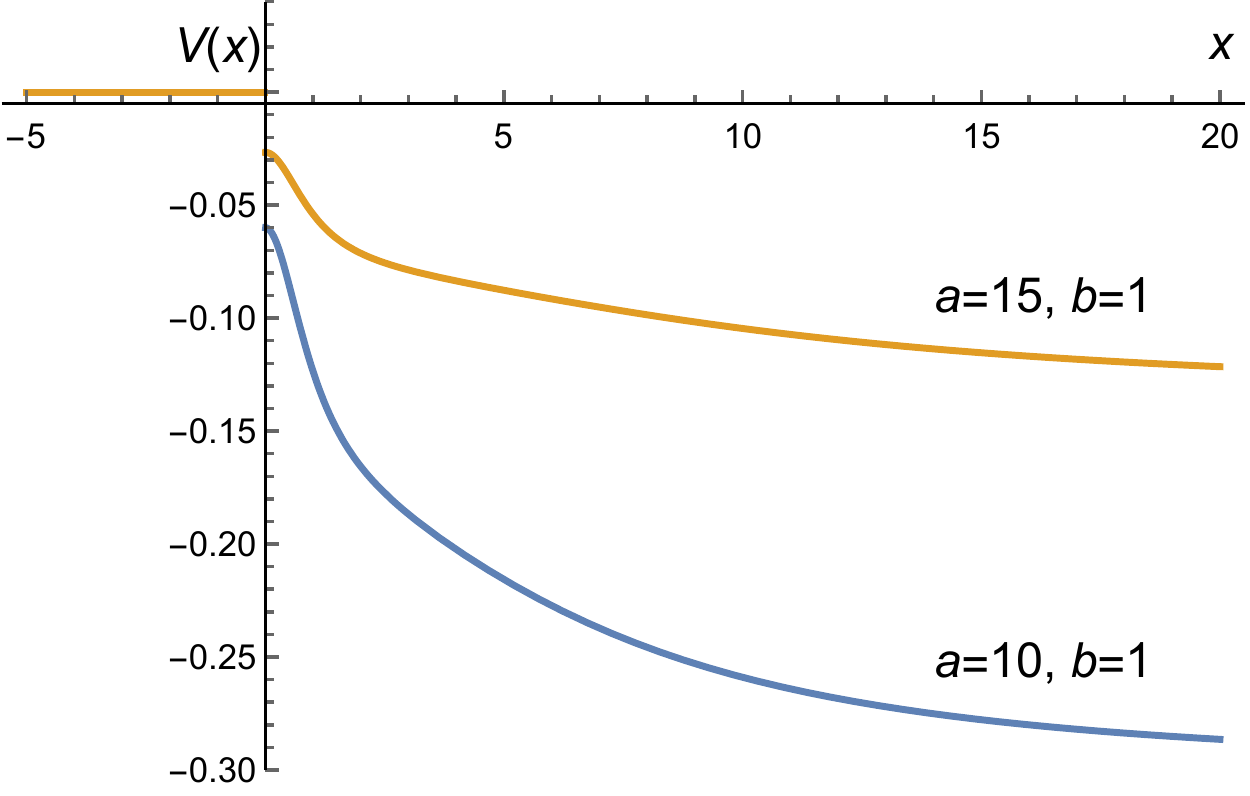}
\caption{\small The scalar field $\phi(x)$ (left) and the potential $V(x)$ (right) in Example 2}
\end{figure*}
  
  On the other hand, the throat radius \rf{r_th} depends on the same constants $n$ and $\ok$
  and will not be too large if they take modest values. Thus, for $n/\ok = 14$, \eqn{r_th}
  gives $r_{\min} \approx 76 \ok p_-$. To obtain a large enough throat for passing of a 
  macroscopic body, say, $r_{\min} = 10$ meters, one has to suppose $\ok \sim 10^{18}$.

\subsection {Example 2: ST asymptotically AdS wormholes}

  With nonzero potentials $V(\phi)$, in most cases solutions can be found 
  only numerically, with one exception: the case $K=0$ in \eqn{05i}. It then follows $P = cA$,
  $c = \const $, and \eqn{02} takes the form
\beq                                        \label{02ii}
                  [A^3 (R/A)']' = 2A.
\eeq
  It is a single equation for two functions $A(x)$ and $R(x)$. It can be solved by 
  quadratures if one specifies $A(x)$: indeed, \eqn{02ii} can be rewritten as
\beq                      \label{R'}
	      \Big(\frac RA\Big)' = \frac {2}{A^3} \int A(x) dx.
\eeq   
  We here want to obtain an example with an \asflat\ 4D space-time on the left end and an
  AdS asymptotic on the right, which corresponds to $A \to 1$ as $x\to -\infty$ and 
  $A \sim x^2$ as $x \to +\infty$. It is, however,  hard to find $A(x)$ with the above 
  properties that would lead to good analytic expressions of other quantities. We therefore 
  consider an example with the following piecewise smooth function $A(x)$:
\bearr             \label{A(x)}    
               A(x) = \vars{1,  &  x \leq 0,\\
                          1 + 3 x^2/a^2, & x\geq 0,   
			}
\ear    
  with $a = \const > 0$. Then we must further solve the equations separately for $x < 0$
  and $x >0$ and match the solutions at $x=0$. At $x < 0$ \eqn{02ii} leads to $R'' = 2$, hence 
  we can take 
\beq                                               \label {R-}
             R(x) \equiv r^2(x) = x^2 + b^2 \quad\  (x \leq 0),
\eeq  
  with $b = \const > 0$ (evidently, this means that $x=0$ will be a throat of radius $b$).
  Further on, from \rf{00} and \rf{01} we find without loss of generality
\beq
	  V(x) \equiv 0, \ \ \   \phi(x) = \arctan(x/b)\ \ \ (x \leq 0).
\eeq 

  The same functions at $x > 0$ are also easily calculated: from \rf{R'} it follows 
\beq
           R(x) = \Big(1 + \frac{3 x^2}{a^2}\Big) \biggl[b^2 
                                    + \frac{x^2 (1 + 2 x^2/a^2)}{(1 + 3 x^2/a^2)^2}\biggr],
\eeq
  where the emerging integration constant is chosen to provide continuity at $x=0$. The 
  potential $V(x)$ is then found from \rf{00} (recall that $P(x) = cA(x)$): 
\bearr  
          V(x) = - \frac{30}{a^2}
\nnn
         + \frac{ 12[b^2 x^2 + a^2 (2b^2 + x^2)] }
          { 9b^2 x^4 + a^4 (b^2 + x^2) + 2a^2 x^2 (3b^2 + x^2) }.
\ear
  At large $x$ ($x\to \infty$), we have 
\bearr              \label {as+}
       R(x) = \Big(\frac 23 + \frac{3 b^2}{a^2}\Big) x^2 + O(1),\quad P(x) = \frac {3cx^2}{a^2}+c, 
\nnn
       V(x) = - \frac{30}{a^2} + \frac {12 (a^2 + b^2)}{(2 a^2 + 9 b^2)x^2} + o(x^{-2}).
\ear
  Thus the solution has a 6D AdS asymptotic behavior (with the curvature radius $a/\sqrt{3}$), 
  corresponding to a negative constant $V$ which here plays the role of a cosmological 
  constant. As already mentioned, the potential $V$ is zero at negative $x$ and has a jump 
  at $x=0$ due to a jump in $A''$. The expression for 
  $\phi(x)$ is found from \rf{01} numerically, see the figure; this function is made continuous 
  at $x=0$ by choosing the integration constants, but the derivative $\phi'$ suffers a jump
  (though it is not evident in the plot). The monotonic nature of $\phi(x)$ makes the potential 
   $V(\phi)$ well defined. The jumps in both $V(x)$ and $\phi'(x)$ at $x=0$ could be easily 
  removed by choosing $A(x)$ smoother than $\rm C^1$ at $x=0$, which is possible by making 
  a suitable arbitrarily small addition to \rf{A(x)}. 

\section{Concluding remarks}

  The results can be summarized as follows. We have considered static solutions with the metric 
  \rf{ds_6} in 6D GR with a minimally coupled scalar field as a source of gravity
  and selected the kinds of asymptotic behavior of the metric functions compatible with the field 
  equations. It has turned out that the choice of possible behaviors is rather narrow, and in particular, 
  Rubin wormholes (as described in the introduction) are impossible in this framework. Instead, 
  we have found another type of wormhole which lead from our universe with small extra
  dimensions to a universe with large extra dimensions where space-time is effectively 6-dimensional
  and should possess quite unusual physics. In our explicit examples of such configurations the extra 
  dimensions have the geometry of a 2-torus. The first example represents a special case of a 
  well-known general solution with a massless scalar field \cite{k95, bim97}, where the extra 
  factor space has a large constant size at the ``far end''; in the other example, with a nonzero potential 
  $V(\phi)$, the ``far end'' has a 6D AdS geometry.

  The existence of such configurations or their analogs with a different number of extra dimensions in
  our universe cannot be a priori excluded, and their possible astrophysical consequences could be a 
  subject of further studies.   

  It should be noted that the analysis performed in Section 3 certainly did not cover all opportunities: 
  we only considered \asflat\ and AdS behaviors of the metric, whereas other, more complicated 
  cases are also possible. For instance, of particular interest is a de Sitter asymptotic which will 
  lead to space-times with horizons and very probably to new cosmological models of ``black universe''
  type, where the cosmological expansion starts from a Killing horizon instead of a singularity, 
  see, e.g. \cite{BR, bu1, bu2, bu3} and references therein.     

  One more subject of a future study can be a relationship between the present scalar-vacuum system 
  and multidimensional gravity with curvature-nonlinear actions \cite{BR, Rub16} in 
  different conformal frames in application to space-times of the type considered here and in 
  \cite{Rub16}.  

\subsection{Acknowledgments} 

 We thank Sergei Rubin and Sergei Bolokhov for numerous helpful discussions. 
 The work of KB was partly performed within the framework of the Center 
  FRPP supported by MEPhI Academic Excellence Project 
  (contract No. 02.a03.21.0005, 27.08.2013).
  This work was also funded by the Ministry of Education and Science of the Russian
  Federation on the program to improve the competitiveness of the RUDN University among the 
  world leading research and education centers in 2016--2020 and by RFBR grant 16-02-00602..

\end{document}